\documentclass[twocolumn, prd, nofootinbib,
showpacs, preprintnumbers, amsmath, amssymb]
              {revtex4}
\usepackage{graphicx}

\begin{document}

\title{Light incoherence due to quantum-gravitational fluctuations of
the background space}

\author{ Michael~Maziashvili}
\email{mishamazia@hotmail.com} \affiliation{ Andronikashvili
Institute of Physics, 6 Tamarashvili St.,
  Tbilisi 0177, Georgia } \affiliation{
Faculty of Physics and Mathematics, Chavchavadze State University,
32 Chavchavadze Ave., Tbilisi 0179, Georgia}

\begin{abstract}

Based on the theory of mutual coherence of light from an extended
incoherent quasi-monochromatic source (providing a basis of
stellar interferometry) we estimate the degree of light
incoherence due to quantum-gravitational fluctuations of the
background metric. It is shown that the stellar interferometry
observational data considered in the literature for a last few
years as a manifestation against the Planck scale
quantum-gravitational fluctuations of the background metric have
no chance for detecting such an effect.

\end{abstract}




\maketitle

\section*{Introduction}

Quantum gravity (QG) strongly indicates the finite resolution of
space-time, that is, space-time uncertainty. A glance at a recent
QG literature obviously shows that space-time uncertainty is
common for all approaches to quantum gravity: space-time
uncertainty relations in string theory \cite{String1, String2};
noncommutative space-time approach \cite{noncommutative}; loop
quantum gravity \cite{Loop}; or space-time uncertainty relations
coming from a simple {\tt Gedankenexperiments} of space-time
measurement \cite{Gedankenexperiments, random, mazia, holographic}. Well known entropy bounds
emerging via the merging of quantum theory and general relativity
also imply finite space-time resolution \cite{entropybounds}. The
combination of quantum theory and general relativity in one or
another way manifests that the conventional notion of distance
breaks down the latest at the Planck scale $l_P \simeq
10^{-33}$\,cm \cite{minimumlength}. Since our understanding of
time is tightly related to the periodic motion along some length
scale, this result implies in general an impossibility of
space\,-\,time distance measurement to a better accuracy than
$\sim l_P$. It is tantamount to say that the space\,-\,time point
undergoes fluctuations of the order of $\sim l_P^4$, that is,
space\,-\,time point is effectively replaced by the cell $\sim
l_P^4$, we refer the reader to a very readable papers of Alden Mead
\cite{Gedankenexperiments} for his discussion regarding the status of a
fundamental (minimum) length $l_P$, as this conceptual standpoint
was unanimous in almost all subsequent papers albeit many authors
apparently did not know that paper. The local fluctuations, $\sim
l_P$, add up over the macroscopic scale $l \gg l_P$ in this or
another way that results in fluctuation $\delta l(l)$. In view of
the fact how the local fluctuations of space\,-\,time add up over
the macroscopic scale different scenarios come into play. Most
interesting in quantum gravity are random and holographic
fluctuations. If the local fluctuations, $\sim l_P$, are of random
nature then over the length scale $l$ they add up as $\delta l =
(l/l_P)^{1/2}l_P$ \cite{random, mazia}. In the holographic case, the local
fluctuations, $\sim l_P$, add up over the length scale $l$ in such
a way to ensure the black hole entropy bound on the horizon region
$\delta l = (l/l_P)^{1/3}l_P$ \cite{holographic, mazia}. In what follows we will consider a
relatively general case by parameterizing the space-time
uncertainty as follows
\begin{equation}\label{lenunc}\delta l = \beta
\,l_P^{\alpha}\,l^{1-\alpha}~,\end{equation} where $1/2 \leq \alpha \leq 1$
and $\beta$ is understood to be of order unity. Eq.(\ref{lenunc})
tells us that because of background metric fluctuations the length
$l$ undergoes fluctuations of the order of $\delta l$. Having
summarized the background metric fluctuations in
Eq.(\ref{lenunc}), let us concisely formulate the problem we want
to address in this paper. Metric fluctuations naturally produce
the fluctuations in energy-momentum of particle, for the particle
with momentum $p$ has the wavelength $\lambda = 2\pi p^{-1}$ and
due to length fluctuation, Eq.(\ref{lenunc}), one finds $\delta p
= 2\pi\lambda^{-2}\delta\lambda,~ \delta E = pE^{-1}\delta p$. An
interesting idea for detecting the space-time fluctuations was
proposed in \cite{LH}. The idea is to consider a phase incoherence
of light coming to us from extragalactic sources. The
energy-momentum uncertainties result in uncertainties of phase and
group velocities of the photon leading to the phase incoherence of
photon during the propagation. Since the phase coherence of light
from an astronomical source incident upon a two-element
interferometer is necessary condition to subsequently form
interference fringes, such observations offer by far the most
sensitive and uncontroversial test. The interference pattern when
the source is viewed through a telescope will be destroyed if the
phase incoherence, $\delta \varphi$, approaches $2\pi$. In other
words, if the light with wavelength $\lambda$ received from a
celestial optical source located at a distance $l$ away produces
the normal interference pattern, the corresponding phase
uncertainty should satisfy the condition $\delta\varphi < 2\pi $.
Soon after the appearance of paper \cite{LH} it was noticed in
\cite{Coule} that such a naive approach overestimates the effect
as the authors of \cite{LH} do not take into account the fact that
the light coming from the distant stellar objects is incoherent
from the beginning but acquires a mutual coherence later simply by
propagation of rays over a large distance with respect to the van
Cittert-Zernike theorem \cite{BornWolf}. In order to estimate the
correct size of the effect, let us follow a textbook
\cite{BornWolf}, which gives the basics concerning the stellar
interferometry.

\section*{Mutual coherence of light from an extended incoherent
  quasi-monochromatic source}

Light from a real physical source is never strictly monochromatic
but
rather
quasi-monochromatic, even the sharpest spectral line has a finite
width. In a
monochromatic wave the amplitude at any point $P$ is constant,
while
the phase
varies linearly with time, that is, a general monochromatic wave
of
frequency
$\omega$ can be represented as a solution of the wave equation of
the
form
\[\varphi(\vec{r},~t) = a(\vec{r})e^{i\left[g(\vec{r})-\omega t\right]
}~. \] This is no longer the case in a wave produced by a
real source: the amplitude and phase undergo irregular
fluctuations,
the
rapidity of which depends on the width of spectrum $\delta\omega$,

\begin{equation}\label{wpacket}\varphi(r,~t) =
\int\limits_0\limits^{\infty}d\omega
\,a(r,~\omega)e^{i\left[k(\omega)\,r-\omega
    t\right] } ~,  \end{equation} where the amplitudes $a(r,~\omega)$ differ
appreciably from zero only within a narrow range around a mean
frequency
$\bar{\omega}$
\[\bar{\omega}-{\delta\omega\over 2} \leq\, \omega\, \leq
\bar{\omega}+{\delta\omega\over 2}~, ~~~~~{\delta\omega\over
\bar{\omega}}\ll 1~. \] Such a quasi-monochromatic wave which is
usually referred to as a wave packet in the physics literature is
characterized with a group and phase velocities \[ v_{p} =
{\bar{\omega}\over \bar{k}}~,~~~~ v_{g} =
\left.\left({d\omega\over dk}\right)\right |_{\bar{k}}~.
\]

\begin{figure}[t]

\includegraphics[width=0.45\textwidth]{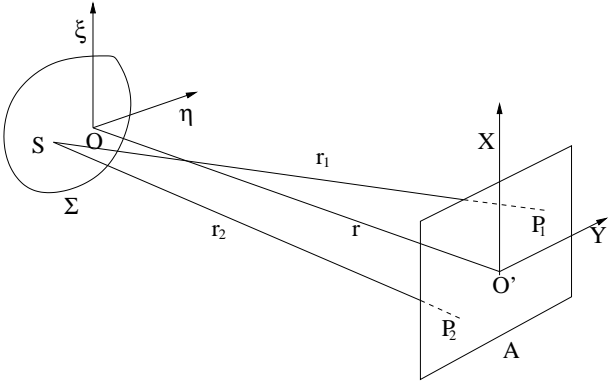}\\
\caption{ A screen $A$ illuminated by an extended
quasi-monochromatic
  incoherent source $\Sigma$. }
\end{figure}

If $\delta\omega$ is sufficiently small the wave packet Eq.(\ref{wpacket}) can be
interpreted as a plane wave with frequency $\bar{\omega}$, wave
number $k(\bar{\omega}) \equiv \bar{k}$ and variable amplitude

\[A(r,~t) = \int\limits_{\bar{\omega}-\delta\omega /2 }\limits^{
\bar{\omega}+\delta\omega / 2}d\omega \,
a(r,~\omega)e^{i\left\{\left[k(\omega) -
      k(\bar{\omega})\right] \,r-\left[\omega - \bar{\omega}\right]
    t\right\}}~. \] The width $\delta\omega$ determines duration of the wave packet
$\delta t \simeq \delta\omega^{-1}$, an important
characteristic for the interference effect. Namely, the interference effect
to take place the path difference between the overlapping quasi-monochromatic
beams must be less than the coherence length $\delta t$. Now let
us follow van Cittert-Zernike approach to the mutual coherence
of light from an extended quasi-monochromatic source. A screen $A$
is illuminated by an extended quasi-monochromatic incoherent
source $\Sigma$ taken for simplicity to be a portion of a plane
parallel to $A$, see Fig.1. The points $P_1$ and $P_2$ correspond to the slits of interferometer.  
We will make the following natural assumptions. The linear dimensions of $\Sigma$ are small
compared to the distance $OO'$ between the source and the screen,
and the angles between $OO'$ and the line joining a typical source
point $S$ to $P_1$ and $P_2$ are small. Dividing the source into
elements $d\sigma_m$ centered on points $S_m$ and denoting
by $\varphi_{m1}(t)$ and $\varphi_{m2}(t)$ the disturbances at
$P_1$ and $P_2$ due to element $d\sigma_m$, for total disturbances
at these points one finds
\[\varphi_{1,2}(t) = \sum\limits_m \varphi_{m1,2}(t)~.\] Correlation
function between the light signals $\varphi_{1}(t)$ and $\varphi_{2}(t)$ takes the form

\[\langle\varphi_1(t)\varphi^*_2(t) \rangle = \sum\limits_m
\langle\varphi_{m1}(t)\varphi^*_{m2}(t) \rangle +
\sum\limits_{m\neq n} \langle\varphi_{m1}(t)\varphi^*_{n2}(t)
\rangle~. \] Because of mutual incoherence, the light signals
coming from different elements of the
source are statistically independent, that is, there is
no correlation between $\varphi_{m1}(t)$ and  $\varphi_{n2}(t)$
when
$m\neq n$
\[\langle\varphi_{m1}(t)\varphi^*_{n2}(t) \rangle =
0~,~~~~\mbox{for}~~m\neq
n~.\] The signals $\varphi_{m1,2}(t)$ represent spherical waves
coming to the points $P_1$ and $P_2$ from the elements $d\sigma_m$
\begin{equation}\label{sphwaves}\varphi_{m1,2}(t) =
A_m(t-r_{m1,2})\,{e^{-i\bar{\omega}[t-r_{m1,2}]}\over
  r_{m1,2} }  ~, \end{equation} where $r_{m1,2}$ denote distances between the
elements $d\sigma_m$ and the points $P_{1,2}$. Using Eq.(\ref{sphwaves}) for the correlation function one finds  
\begin{widetext}
\begin{equation}\label{intcond} \langle\varphi_{m1}(t)\varphi^*_{m2}(t)
\rangle   =  \left\langle A_m(t-r_{m1})A^*_m(t-r_{m2})
\right\rangle \,{e^{i\bar{\omega}(r_{m1}-r_{m2})}\over
r_{m1}\,r_{m2}}
  =  \left\langle A_m(t)A^*_m(t - r_{m2} + r_{m1}) \right\rangle
\,{e^{i\bar{\omega}(r_{m1}-r_{m2})}\over r_{m1}\,r_{m2}}~.
\end{equation}\end{widetext} If the path difference $r_{m2} - r_{m1}$
is small compared to
the coherence length (duration of the wave packet $\delta t$) one
can
neglect
the retardation $r_{m2} - r_{m1}$ in the argument of $A^*$ in
Eq.(\ref{intcond}) that gives
\begin{equation}\label{corfunc}\langle\varphi_1(t)\varphi^*_2(t) \rangle = \sum\limits_m
\left\langle
  A_m(t)A^*_m(t) \right\rangle
  \,{e^{i\bar{\omega}(r_{m1}-r_{m2})}\over
  r_{m1}\,r_{m2}}~. \end{equation} The quantity $\langle
  A_m(t)A^*_m(t)\rangle$ characterizes the intensity of the radiation
from the
  source element $d\sigma_m$. So the correlation function Eq.(\ref{corfunc}) takes
the form \begin{equation}\label{corfunc1} \langle\varphi_1(t)\varphi^*_2(t) \rangle =
  \int\limits_{\Sigma} d\sigma I(\sigma)
\,{e^{i\bar{\omega}(r_{1}-r_{2})}\over
  r_{1}\,r_{2}}~, \end{equation} where $I(s_m)d\sigma_m =
\langle
  A_m(t)A^*_m(t)\rangle $. In most applications the intensity
$I(\sigma)$ may be
  assumed to be uniform on $\Sigma$. To work out the integral Eq.(\ref{corfunc1}) let us
denote by
  $(\xi,~\eta)$ the coordinates of a point $S$, referred to axes at
$O$, and
  let $(x_1,~y_1)$ and $(x_2,~y_2)$ be the coordinates of $P_1$ and
$P_2$
  referred to parallel axes at $O'$, see Fig.1. Retaining only leading terms in
$x/r,~y/r,~\xi/r,~\eta/r$, where $r$ is the distance $OO'$, one finds
  \[r_1 - r_2  \approx  { x_1^2 + y_1^2 - x_2^2 - y_2^2 + 2\xi (x_2 -
    x_1) + 2\eta (y_2 - y_1) \over 2r} ~.\] Denoting \[p = {x_1 - x_2
\over r
  }~,~q = {y_1 - y_2 \over r}~,~\psi = {\bar{\omega} (x_1^2 + y_1^2 -
x_2^2 -
    y_2^2) \over 2r }~,\] the Eq.(\ref{corfunc1}) takes the form
\begin{equation}\label{corfunc2}\langle\varphi_1(t)\varphi^*_2(t) \rangle \approx
{e^{i\psi}\over r^2}
  \int\limits_{\Sigma} d\xi d\eta I(\xi,~\eta) \,e^{i\bar{\omega}(p\xi
-
    q\eta)}~. \end{equation} For a uniform circular source of
radius $\rho$
with its center at $O$, the Eq.(\ref{corfunc2}) reduces to

  \begin{equation}\label{mutcoh} \langle\varphi_1(t)\varphi^*_2(t)
\rangle \sim  e^{i\psi}{J_1(v)\over
    v}~, \end{equation} where $v=\bar{\omega}\rho\sqrt{p^2 + q^2} $ and
$J_1$ stands for
  the Bessel function of the first kind and first order. In most
applications
  the quantity $\psi$ is very small, so that one can neglect
corresponding phase factor
  in Eq.(\ref{mutcoh}). The function
  $J_1(v)/v$ decreases steadily from the value $0.5$ when $v=0$ to the
value
  zero when $v=3.83$ indicating that the degree of coherence steadily
  decreases and approaches complete incoherence when $P_1$ and $P_2$
are
  separated by the distance \[P_1P_2 = {0.61\, \bar{\lambda}\, r \over
    \rho}~.\] In experiments on interference and diffraction a
departure of
  $12$ per cent from the ideal value of coherence that occurs at $v=1$
can be
  taken as a maximum permissible departure that gives for the
separation of
  points $ P_1$ and $P_2$ \begin{equation}\label{explim} P_1P_2 =
{0.16\, \bar{\lambda}\, r \over\rho}~.\end{equation}

\section*{The degree of incoherence due to QG fluctuations}

As it was said in the introduction, due to quantum-gravitational
fluctuations of the background metric the length scale $l$
undergoes fluctuations of the order of $ \delta l$
Eq.(\ref{lenunc}). Because of background metric fluctuations there
is an increment of the wave packet width which can be simply
estimated by using the Eq.(\ref{lenunc}) \begin{equation}\label{width}\delta \omega =
\bar{\omega}\, {\delta\bar{\lambda} \over \bar{\lambda}} =
\bar{\omega}\,\beta\left({l_P\over\bar{\lambda}}\right)^{\alpha}~.
\end{equation}  A wavelength of the light from stellar objects considered in
\cite{LH, RTG, Stein} is in the region $\bar{\lambda} \simeq \mu$m
and correspondingly for the width increment of a wave packet from Eq.(\ref{width}) one
finds ($1/2 \leq \alpha \leq 1$ and $\beta$ is of order unity) \begin{equation}\label{width1} {\delta \omega \over \bar{\omega}} \simeq \beta\,
10^{-29\alpha}\ll 1 ~.\end{equation} In addition we need to
check an important requirement that the path difference
$|r_{m2}-r_{m1}|$ is still smaller than $\delta\omega^{-1}$. We
have $|r_{m2}-r_{m1}| \leq P_1P_2$, where with respect to the
observational data considered in \cite{LH, RTG, Stein}, $P_1P_2$
is in the range $1$m - $20$m (the distance between two slits of the
interferometer). Recalling that $1/2 \leq \alpha \leq 1$ and $\beta$ takes on reasonable values, for the wavelength $\bar{\lambda} \simeq \mu$m from Eq.(\ref{width1}) one finds \[\delta \omega^{-1} \simeq
\bar{\lambda}\,{10^{29\alpha} \over \beta} \gg 20\mbox{m}~,
~.\] So one infers that there is no
change in the coherence picture due to fluctuations of
$\bar{\lambda}$. The observational data considered in \cite{LH,
RTG, Stein} are collected from high-redshift astronomical sources
with maximum $z = 5.4$. So that the photons left the sources at
the shorter wavelength $\lambda /(1 + z) $. But it is easy to see
that the above conclusions do not change by inclusion the factor
$z = 5.4$.

In light of the observational data considered in \cite{LH, RTG,
Stein}, the maximum value of $\psi$ reads \[\psi =
{\pi \left(P_1P_2\right)^2\over
  \bar{\lambda}\, r } \simeq 4\cdot 10^{-11}~,  \] and corresponding
phase
factor in Eq.(\ref{mutcoh}) may evidently be neglected.

Now by taking the variations of $\rho,~r$ in Eq.(\ref{explim}) one
finds
\begin{equation}\label{effrate} \delta\left(P_1P_2\right) \simeq  \beta
\left( P_1P_2 \right)^{1+\alpha} \left({l_P \over 0.16\cdot
    \bar{\lambda} \cdot r} \right)^{\alpha}~. \end{equation} Let us
estimate
the maximum of this variation by choosing the corresponding
parameters
from
the data \cite{LH, RTG, Stein}, that is, $r \sim 1$kpc, $P_1P_2 \sim
10^3$cm,
$\bar{\lambda} \sim 10^{-4}$cm. For this set of parameters from
Eq.(\ref{effrate}) one finds
\begin{equation}\label{maxrate} \delta\left(P_1P_2\right) \sim \beta\,
  10^{3-47\alpha}\mbox{cm}~,  \end{equation} which for $\alpha = 1/2$
(that gives a maximum as $1/2 \leq \alpha \leq 1$) yields

\begin{equation}\label{maxalphrate}  \delta\left(P_1P_2\right) \sim
\beta\, 10^{-20}\mbox{cm}~. \end{equation} So that the
observations analyzed in \cite{LH, RTG, Stein} have no chance to
detect the quantum gravitational fluctuations described by the
Eq.(\ref{lenunc}) for any reasonable value of $\beta$ and $1/2
\leq \alpha \leq 1$. The effect to be somewhat appreciable the
ratio $\delta\left(P_1P_2\right)/ P_1P_2 $ should not be highly
suppressed, that is, it should be possible to control the length
$P_1P_2 $ with the accuracy $\delta\left(P_1P_2\right)$.

\section*{ Concluding remarks }

First let us outline how the problem of light incoherence from
distant astronomical sources caused by the QG fluctuations of the
background metric was treated in the previous study. In paper \cite{LH} the authors assumed that the light
coming from the distant extragalactic sources (the
diffraction/interference images of which are seen through the two
slit telescopes) is coherent from the beginning, but accumulates
appreciable phase incoherence if the length of propagation, $t$,
is large enough. Phase incoherence accumulates at the expense of factor $t\delta\omega$ and becomes sufficiently large even for
small values of $\delta\omega$ ($\delta\omega$ is caused by the QG
fluctuations of the background metric) when $t$ is large enough. So it is understood
that the time dependance of the wave, $t\omega$, varies due to QG
fluctuations as $\delta (t\omega) = \omega\delta t +
t\delta\omega$ and because the second term is dominating it is
taken as a main source of phase incoherence. The condition
$t\delta\omega \geq 2\pi$ is understood as a criterion for
incoherence that should lead to the destruction of the
diffraction/interference patterns when the source is viewed
through a telescope. From Eq.(\ref{width1}) one finds $\delta \omega$ for a given $\bar{\omega}$ and then readily estimates (for a given value of $\alpha$) the distance $t \simeq 2\pi / \delta\omega$. 

Following this way of reasoning, in paper \cite{RTG} the distance through which the
wave-front recedes when the phase increases by $t\delta\omega$ is
taken as an error in measurement of a length, $t$, by the light
with wavelength $2\pi/\omega$, and due to this length variation an
apparent blurring of distant point sources is estimated. Those
papers found the amplification of Planck scale 
tiny QG effect to an extent that contradicts the existing observations of
diffraction/interference images through the telescopes.

In order to mitigate the situation it was suggested in \cite{NgvDC} that
the phase incoherence
\begin{equation}\label{cumfac}t\delta\omega = \omega\,{t\over
\lambda}\,\delta\lambda~,\end{equation} could be reduced by taking
the factor $(l/\lambda)^{1-\alpha}$ instead of $l/\lambda$ as it
stands in Eq.(\ref{cumfac}). For motivating this step, the authors
of paper \cite{NgvDC} were assuming that in the case $\alpha =1/2$ the
fluctuations $\delta\lambda$ were expected to take on $\pm$ sign
with equal probability and respectively had to add up over the
length scale $t$ by the cumulative factor $(l/\lambda)^{1/2}$.
Then for other values of $\alpha$, the cumulative factor was
assumed of the form $(l/\lambda)^{1-\alpha}$. (This reduced
expression for the phase incoherence is used in \cite{Stein} as
well.) First, one may question that as in the particular case $\alpha
=1/2$ the local fluctuations $\sim l_P$ add up over the
macroscopic scale $l$ to $\delta l = \beta (l/l_P)^{1/2}l_P$ that
obviously contradicts the argument that $\delta\lambda$ should also add
by the same cumulative factor. Second, what is most important,
this artificial trick has nothing to do with the physics behind
the phenomenon under consideration. As we see the approaches of these papers to 
the problem of stellar interferometry are oversimplified that leads to the misleading results. The ideology of papers
\cite{LH, RTG, Stein, NgvDC} has little to do with the theory of
mutual coherence of light from an extended incoherent
quasi-monochromatic source, that may be the only reliable way for
a proper understanding of question under consideration.

The discussion of our paper is based on the theory of mutual
coherence of light from an extended incoherent quasi-monochromatic
source (which provides the basis of stellar interferometry). What
we learn from this theory is that the light coming from a real
astronomical source has a natural finite width $\delta \omega$
from the very outset and the contribution to the width due to QG
fluctuations is negligible. Moreover, even were it not so this
would not affect the mutual coherence of light as long as $\delta
\omega$ satisfies the conditions \[ {\delta\omega\over
\bar{\omega}}\ll 1~,~~~~~~|r_{m1} - r_{m2}| \ll
\delta\omega^{-1}~. \] The rate
of QG incoherence is discouragingly small as it is demonstrated by
the Eqs.(\ref{maxrate},\,\ref{maxalphrate}), to be detectable by
the stellar interferometry observations considered in \cite{LH,
RTG, Stein, NgvDC}.

\begin{acknowledgments}

Author is indebted to Professors Jean-Marie Fr\`{e}re and Peter
Tinyakov for invitation and hospitality at the \emph{ Service de
Physique Th\'eorique,
  Universit\'e Libre de Bruxelles}, where a great deal of this paper was
  done. Special thanks are due to Peter
Tinyakov for many useful discussions and comments. The work was
supported in part by the \emph{INTAS Fellowship for Young
Scientists}; \emph{CRDF/GRDF} grant and the \emph{Georgian
President Fellowship for Young Scientists}.

\end{acknowledgments}

\end{document}